\documentclass[a4paper,11pt]{article}
\usepackage{pos}

\title{Extractions of the strong coupling from collider data without PDF refitting are biased}

\author[a]{Stefano Forte}
\author[b]{Juan Rojo}
\author*[c]{Roy Stegeman}
\affiliation[a]{Tif Lab, Dipartimento di Fisica, Universit\`a di Milano and\\
      INFN, Sezione di Milano, Via Celoria 16, I-20133 Milano, Italy}
\affiliation[b]{Department of Physics and Astronomy, Vrije Universiteit, NL-1081 HV Amsterdam and \\ Nikhef Theory Group, Science Park 105, 1098 XG Amsterdam, The Netherlands}
\affiliation[c]{The Higgs Centre for Theoretical Physics, University of Edinburgh,
      JCMB, KB, Mayfield Rd, Edinburgh EH9 3JZ, Scotland}

\emailAdd{r.stegeman@ed.ac.uk}

\abstract{We present an explicit demonstration that a determination  of the strong coupling constant $\alpha_s(m_Z)$ from deep-inelastic scattering and hadron collider data without a simultaneous determination of the parton distribution functions (PDFs) leads to a biased result for both the central value and the uncertainty, even in the ideal scenario (closure test) where there are no internal tensions between datasets and where theoretical calculations describe perfectly the experimental measurements.
Specifically, we show that a determination of $\alpha_s(m_Z)$ from a single process leads in general to a result that differs from the global best fit more than the value of $\alpha_s(m_Z)$ that is actually favoured by this process.

}

\FullConference{The European Physical Society Conference on High Energy Physics (EPS-HEP2025)\\
7-11 July 2025\\
Marseille, France\\}

\begin{document}
\maketitle

The precise knowledge of the strong coupling  $\alpha_s(m_Z)$ and its running with the scale is one of the main bottlenecks towards reaching percent or
sub-percent accuracy in the computation of hadron collider
processes~\cite{Salam:2017qdl}.
Conversely, many of the most accurate
determinations of the strong coupling are obtained from processes that
involve hadrons in the initial
state~\cite{dEnterria:2022hzv,ParticleDataGroup:2024cfk}. These
determinations inevitably involve knowledge of  hadron structure,
as encoded in the parton distribution functions (PDFs) of the proton~\cite{Gao:2017yyd,Kovarik:2019xvh}.
It has now been known for some time~\cite{Forte:2020pyp} that  reliable unbiased results can only be obtained if $\alpha_s$  and the PDFs are simultaneously determined, as opposed to extracting $\alpha_s$ using a fixed PDF set.

Several PDF fitting collaboration have performed simultaneous determinations form $\alpha_s$ and the PDFs for many years.
Specially since the publication of Ref.~\cite{Forte:2020pyp}, analyses carried out outside the PDF fitting collaborations, for example from the LHC experiments, have accounted for the correlations between PDFs and $\alpha_s$ by profiling the PDFs~\cite{ATLAS:2023lhg}.
In this procedure, the PDF eigenvectors are treated as a nuisance parameters that are refitted.
PDF profiling assesses the impact of fitting the PDFs to a new dataset assuming that variations with respect to the baseline PDF set are moderate.

Recently, the NNPDF collaboration carried out an $\alpha_s(m_Z)$ extraction at NLO in the QED and N$^3$LO in the QCD expansion~\cite{Ball:2025xgq} based on  the NNPDF4.0 global fit~\cite{Ball:2021leu,NNPDF:2024djq,NNPDF:2024nan,NNPDF:2024dpb},
which is the determination of $\alpha_s$ from PDFs at the highest perturbative accuracy to date. This determination went beyond the state-of-the-art in several other ways as well: 1) it was the first determination where missing higher order uncertainties (MHOUs) estimated from scale variations were systematically included, 2) it was performed using both frequentist Monte Carlo sampling and Bayesian inference, and 3) the methodology was tested in a closure test where the underlying truth is assumed to be known. The final result found in~\cite{Ball:2025xgq} was
\begin{equation}
\label{eq:alphaval}    \alpha_s(m_Z)=0.1194^{+0.0007}_{-0.0014} \quad \text{ at } \mathrm{aN}^3\mathrm{LO}_{\rm QCD}\otimes {\rm NLO}_{\rm QED} \text{ accuracy}.
\end{equation}

These results were thoroughly validated by closure tests.
In a closure test~\cite{NNPDF:2014otw}, true experimental data are replaced by artificial data, produced adding to  theory predictions obtained with a known fixed true underlying PDF a level of noise obtained  by sampling the experimental covariance matrix. Hence, PDFs fitted to the artificial data are obtained by construction with perfect theory and  faithful experimental uncertainties. Comparing them to the known true underlying PDF allows one to test whether their uncertainties are also faithful. Similarly, if one fits both $\alpha_s$ and the PDFs to the artificial data, one can test whether the value of the strong coupling and its uncertainty are consistent with the known value used to generate the data in the first place. The power of the closure tests resides in the fact that it can be repeated, by regenerating the artificial data many times. This corresponds to "runs of the universe", thereby allowing one to test that the distribution of results displays fluctuations that are consistent with the nominal uncertainty.  Hence in Ref.~\cite{Ball:2025xgq} it was possible to test that the distribution of values of $\alpha_s$ obtained in a population of closure tests was consistent with the known true central value with its nominal uncertainty.

\begin{figure}[t]
    \centering
\includegraphics[width=0.8\textwidth]{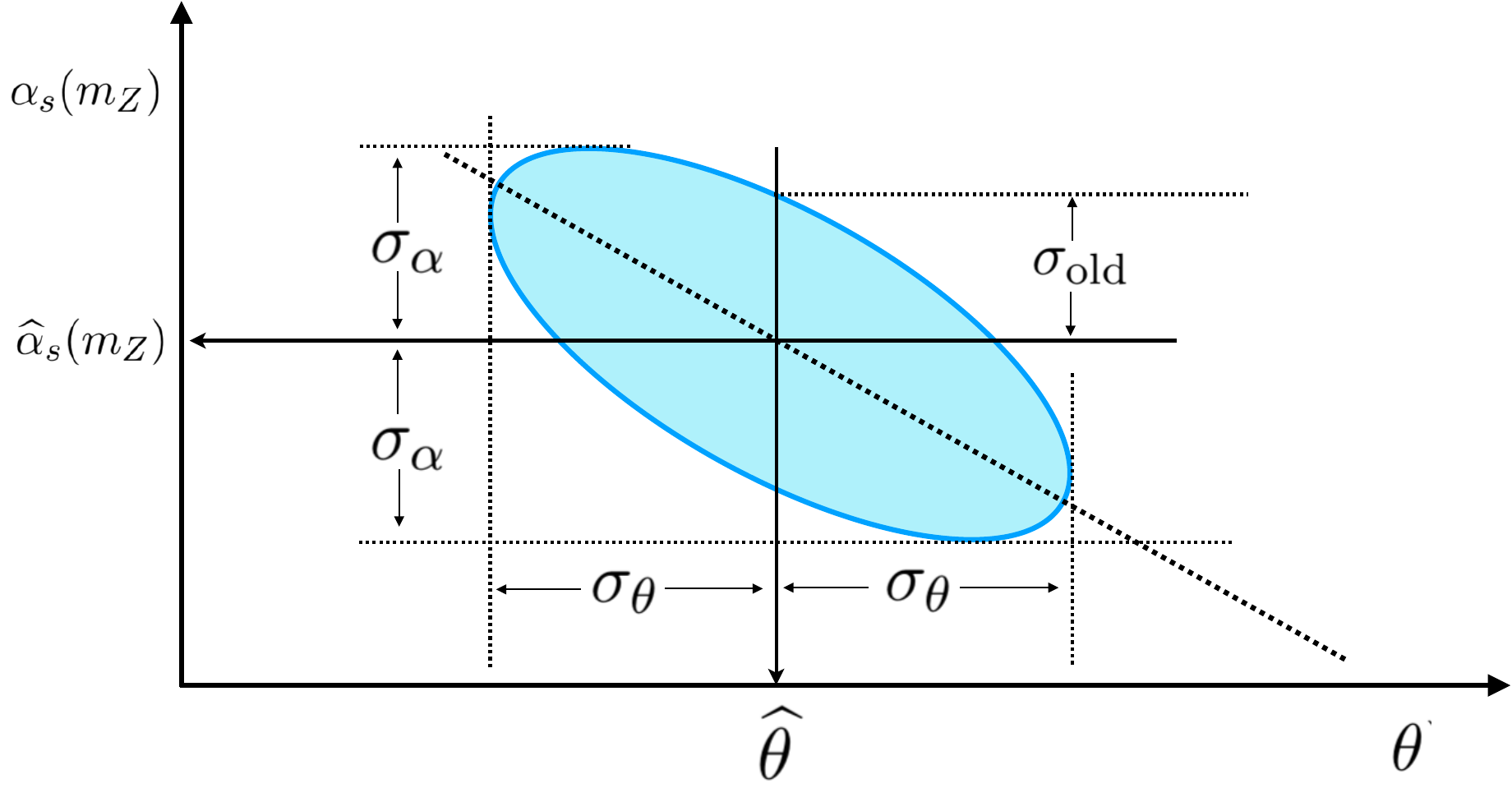}
    \caption{Comparison between the standard deviation of a pair of correlated variables $(\alpha_s,\theta)$ and the one-sigma range for the variable $\alpha_s$ along
the best-fit line of $\theta$. The true uncertainty on $\alpha_s$ is denoted by $\sigma_\alpha$ and the uncertainty obtained by not simultaneously fitting $\theta$ is denoted by $\sigma_{\rm old}$. (Figure taken from Ref.~\cite{Ball:2018iqk}). }
    \label{fig:oldalpha}
\end{figure}

We can use this powerful methodology to provide an explicit demonstration of the formal arguments of Ref.~\cite{Forte:2020pyp}, namely that the extraction of $\alpha_s$ from DIS and collider data without a simultaneous determination of the PDFs leads to biased results. A first simple observation, already made in Ref.~\cite{Ball:2018iqk}, is that determining $\alpha_s$ without simultaneously determining the PDF leads in general to an underestimation of the uncertainty because it neglects the correlation between PDF and $\alpha_s$ uncertainties. This is illustrated in Fig.~\ref{fig:oldalpha}, where $\theta$ stands for the collective set of PDF parameters. If $\theta$ is not fitted and the uncertainty on $\alpha_s$ is determined as the
one-sigma range for $\alpha_s$ with $\theta$ fixed at its best fit value then the resulting uncertainty is $\sigma_{\rm old}$, while the true correlated uncertainty is $\sigma_\alpha$. Indeed if the extraction of $\alpha_s$ of Ref.~\cite{Ball:2025xgq} were performed in this way, the value of Eq.~(\ref{eq:alphaval}) would become $\alpha_s(m_Z)=0.1193^{+0.0005}_{-0.0012}$. Note that the lower uncertainty is obtained by adding to the fit uncertainty a fixed bias, so this means that the fit uncertainty $\sigma_\alpha=0.0007$ would be underestimated to be $\sigma_{\rm old}=0.0005$, which is quite significant.

\begin{figure}[t]
    \centering
       \includegraphics[width=0.49\textwidth]{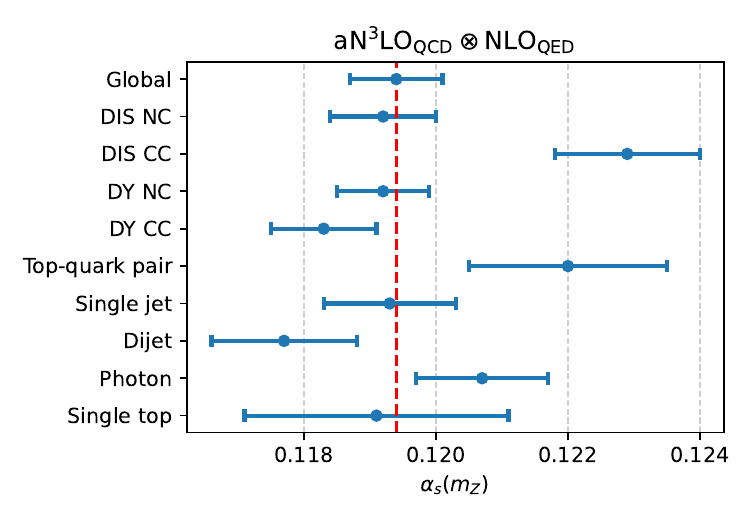}
\includegraphics[width=0.49\textwidth]{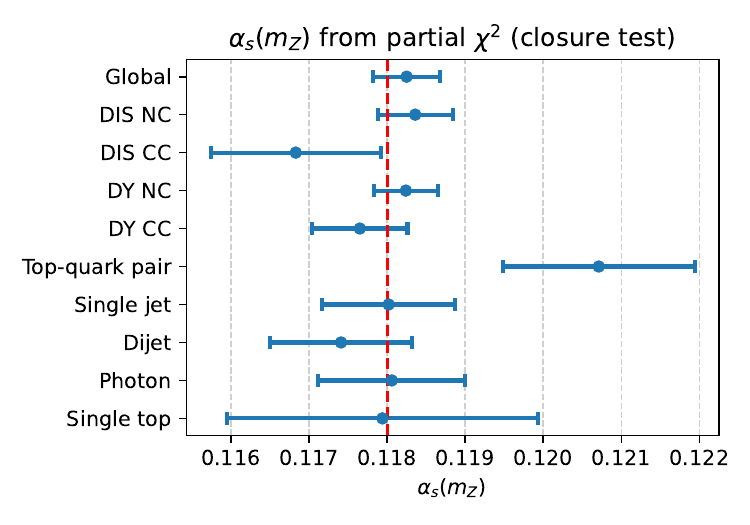}
   \caption{Left: values of $\alpha_s(m_z)$ extracted from the partial $\chi^2$ (real data) evaluated for separate groups of processes for the determination of Ref.~\cite{Ball:2025xgq}.
   The vertical dashed line indicates the best fit value from the global fit.
   Right: same, now for the $\alpha_s(m_z)$ values extracted from closure test artificial data generated with $\alpha_s^{\rm true}=0.118$ (vertical dashed line).
    }
    \label{fig:procs}
\end{figure}

However, the issue exposed in Ref.~\cite{Forte:2020pyp}, which we now demonstrate in a closure test, is rather subtler.
The argument presented in Ref.~\cite{Forte:2020pyp} starts from the observation  that performing a simultaneous determination of the strong coupling constant and the PDFs from different (sub)sets of data generally gives different results because the data fluctuate within their uncertainties. This is illustrated in Fig.~\ref{fig:procs} (left), where we show the partial values of $\alpha_s$ obtained in Ref.~\cite{Ball:2025xgq}, and corresponding to the central value given in Eq.~(\ref{eq:alphaval}). The value denoted as "global" in the figure is that of
Eq.~(\ref{eq:alphaval}), but without the aforementioned contribution to the downward uncertainty, so $\sigma_\alpha=0.0007$. As mentioned, this result was obtained in Ref.~\cite{Ball:2025xgq} using both frequentist Monte Carlo sampling and Bayesian inference.
In both cases the quantity that determines $\alpha_s$ is the $\chi^2$ computed from the data, including a covariance matrix that accounts for fully correlated experimental uncertainties and theoretical missing higher-order uncertainties. The partial $\alpha_s$ values shown are obtained by only using the $\chi^2$ evaluated from a subset of data, but with the same PDFs as in the global determination.
The fact that these partial $\alpha_s$ display wide fluctuations was already observed in Ref.~\cite{Ball:2018iqk}, where a similar behaviour was observed, and motivated the study of Ref.~\cite{Forte:2020pyp}.

In Fig.~\ref{fig:procs} (right) we also show the global and partial $\alpha_s$ values, but now obtained in a closure test, with artificial data generated assuming $\alpha_s(m_Z)=0.118$.
The fact that top-quark pair production data are  showing a similar deviation in the closure test and in the real data is, of course, an entirely accidental feature of this particular set of closure test data,  and is not true with other choices of closure test data. Indeed we have picked a closure test situation in which the pattern is similar to that of real data in order to show that this is a generic situation. Because the  data and theory in the closure test are perfect, the fluctuations seen in this case cannot be due to putative data inconsistency. They are necessarily purely due to statistical fluctuations of the underlying data, amplified by the effect of  Ref.~\cite{Forte:2020pyp}, which we now explain.

\begin{figure}[t]
    \centering
\includegraphics[width=0.8\textwidth]{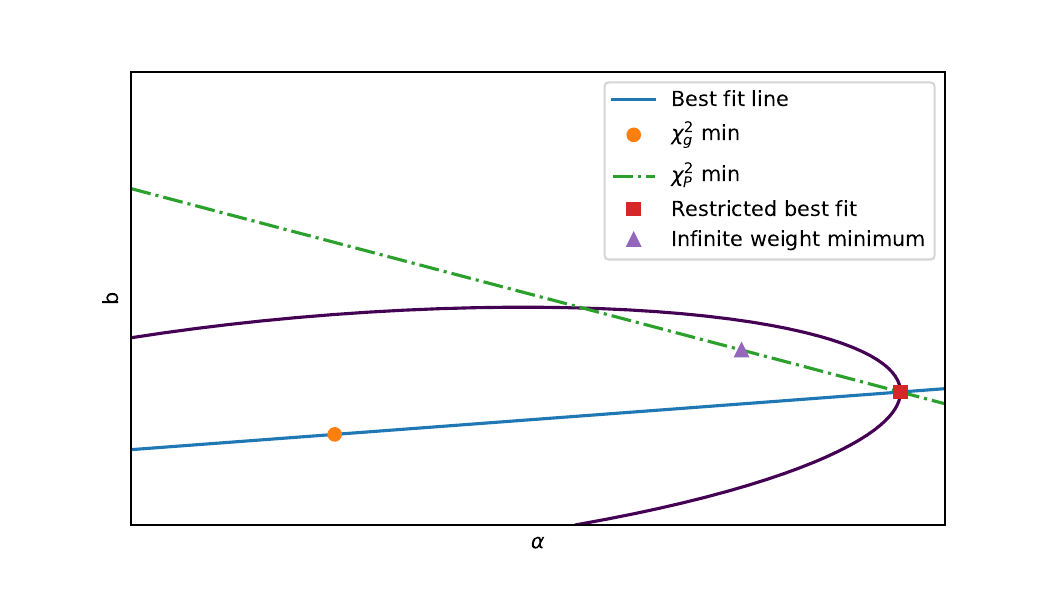}
    \caption{Contours of $\chi^2$ contours in (PDF, $\alpha_s$) space (see text)
(Figure taken from Ref.~\cite{Forte:2020pyp}).
    }
    \label{fig:fortekassabov}
\end{figure}

The origin of the  effect is represented schematically in Fig.~\ref{fig:fortekassabov}. In this figure, similarly to  Fig.~\ref{fig:oldalpha}, the $\chi^2$ contours are shown in a plane where one axis corresponds to values of $\alpha_s$, and the other to a parameter, here called $b$,
analogous to $\theta$ of Fig.~\ref{fig:oldalpha}, that collectively represent the PDFs.  As in Fig~\ref{fig:oldalpha}, the ellipse is a contour of fixed global $\chi^2$ in ($\alpha_s$, PDF) space, with $\chi^2_g$ representing the global best fit.  The best-fit line corresponds to the global best-fit PDF for each value of $\alpha_s$. The $\chi^2_p$~min line corresponds to the locus of points  that provide the best-fit to a subset of data: such as for instance top data. It is assumed that  the dataset does not fully determine the PDFs (like top data, which determine almost only the gluon~\cite{Czakon:2016olj}), so it is a curve in ($\alpha_s$, PDF space), drawn as a straight line under the assumption that one is looking at small deviations.

The observation of Ref.~\cite{Forte:2020pyp} is that the value of $\alpha_s$ obtained by looking at the minimum of the $\chi^2$ for the subset of data  if one does not refit the PDFs is the intersection of the best-fit line and the $\chi^2_p$~min line: indeed, it is simultaneously the best-fit PDF for each $\alpha_s$ value (so it is along the best-fit line), and the best-fit  for the given data subset (so it is along the $\chi^2_p$~min line). This point, denoted as restricted best fit in the figure, can be relatively far from the global minimum because of statistical fluctuations: for example in the case of the top  data of Fig.~\ref{fig:procs} it is about two sigma away, meaning that this point is located on the two sigma ellipse.
However, it is clear that any point along the $\chi^2_p$~min line that  lies within the ellipse provides an equally good fit to the data subset, but a better global fit. In fact, if one were to perform a fit that includes this dataset, but with a large enough weight given to it, then this weighted {\it global} best-fit would be along the $\chi^2_p$~min, but at the point where also the $\chi^2$ of all other data is minimum, which is depicted as "infinite weight minimum" in the figure. This corresponds to a value of $\alpha_s$ which is actually closer to the global minimum than the restricted best fit. Hence, the deviation of the restricted best fit from the global best fit is actually larger than that of an equally good fit to the partial dataset, but a better fit to the global dataset. It was shown in Ref.~\cite{Forte:2020pyp} that similar conclusions apply to the case of a data subset (such as deep-inelastic scattering) that does also fully determine the PDFs.

We can now explicitly test this scenario in the closure test. We start with the fit shown in
 Fig.~\ref{fig:procs} (right), and we note that the partial $\alpha_s$ value obtained from the top data is
\begin{equation}\label{eq:partialtop}
 \alpha_s^{t\bar{t}}(m_Z)=0.1207\pm0.0012,
 \end{equation}
 while the global value is
\begin{equation}\label{eq:globalclos}
 \alpha_s^{\rm global}(m_Z)=0.1183\pm0.0004 \, .
 \end{equation}
 We then perform a new global PDF determination in which the top quark pair production data are now reweighted by a factor $N^{\rm tot}_{\rm dat}/N^{t\bar{t}}_{\rm dat}$, i.e. such that they carry the same weight in the total $\chi^2$ as all the other data combined. In this case we find
\begin{equation}\label{eq:partialtopw}
 \alpha_s^{t\bar{t}}(m_Z)=0.1201\pm0.0012 \, .
 \end{equation}
This is a one-and-a-half sigma shift in units of the uncertainty on the global value of Eq.~(\ref{eq:globalclos}).  In fact, whereas the partial value Eq.~(\ref{eq:partialtop}), differs by 2.3 sigma from the global value (in units of its own uncertainty), the weighted value
Eq.~(\ref{eq:partialtopw}) only differs 1.8 sigma.
What could be have been taken as a sign of tension between the top data and the rest is now identified as a statistical fluctuation --- as we know it to be, in the context of the closure test, where it is by construction.

Based on this exercise, one might argue that the value of $\alpha_s$ that is preferred by each dataset included in a global fit is the value obtained by giving a large weight to each dataset in turn. Indeed, this is the value that, with suitably refitted PDFs, provides the best fit to the given dataset, while also providing the best possible fit to all other data. Be that as it may, the important conclusion is that there generally exist points in ($\alpha_s$, PDF) space that provide a better fit to the global dataset while providing an equally good fit to partial datasets, that correspond generally to a value of $\alpha_s$ that is closer to the global value than the naive partial $\alpha_s$, obtained by looking at the minimum for a dataset along the global best-fit like, i.e. without refitting the PDFs. Hence, values of $\alpha_s$ obtained from partial datasets without refitting the PDFs will generally display inflated fluctuations with respect to the global value, which may suggest tensions between data when there are in fact only statistical fluctuations. More importantly, combining such partial $\alpha_s$ values as if they were independent determinations  leads to a biased combination as it ignores their underlying correlation through the PDF, whose neglect  is responsible for the bias that we have exposed.

\bibliographystyle{utphys}
\bibliography{epshep_alphas}

\end{document}